\documentclass[12pt,preprint]{aastex}
\usepackage{epsfig}
\pdfoutput=1
\usepackage{natbib}
\begin{document}

\title{INTRINSIC SHAPES OF ELLIPTICAL GALAXIES}

\author{DAVID MERRITT}

\begin{abstract}
Tests  for  the intrinsic  shape  of  the  luminosity distribution  in
elliptical   galaxies  are   discussed,  with   an  emphasis   on  the
uncertainties.  Recent  determinations  of the  ellipticity  frequency
function  imply a  paucity of  nearly spherical  galaxies, and  may be
inconsistent with  the oblate hypothesis.  Statistical  tests based on
the correlation  of surface brightness, isophotal  twisting, and minor
axis  rotation  with  ellipticity  have  so far  not  provided  strong
evidence in favor  of the nearly oblate or  nearly prolate hypothesis,
but  are  at  least  qualitatively consistent  with  triaxiality.  The
possibility  that   the  observed  deviations   of  elliptical  galaxy
isophotes   form   ellipses  are   due   to   projection  effects   is
evaluated.  Dynamical   instabilities  may  explain   the  absence  of
elliptical galaxies flatter than about E6,  and my also play a role in
the lack of nearly-spherical galaxies.
\end{abstract}

\section{Introduction}

About ten  years ago, a  number of observational programs  designed to
elucidate  the   systematics  of  elliptical   galaxy  morphology  and
kinematics were begun.   The new data generated a  flurry of papers in
the  early  1980s  on  the  subject  of  elliptical  galaxy  intrinsic
shapes.  The question  was usually  phrased  in one  of two  ways:
either ``Are elliptical galaxies  more nearly oblate or prolate?,'' or
``Is there strong evidence for triaxiality?''. While the answer to the
second question was generally felt  to be ``yes'', none of the studies
provided convincing evidence  of a preference in nature  for oblate or
prolate forms (perhaps because  most elliptical galaxies are maximally
triaxial). After a  hiatus of several years, the  topic has once again
begun to attract  attention; one of the new  developments has been the
widespread  use  of CCDs,  which  have  permitted  much more  accurate
measurements of isophotal shapes than in the past. Rather than attempt
a comprehensive review of the subject -- Schecter's summary of 1987 is
still essentially up  to date -- I will focus  on the uncertainties in
this  game, and  on  possible alternative  interpretations. While  the
problem of intrinsic shapes is a  difficult one, the
accumulation of high  quality data over the last few  years puts us in
an excellent position to improve on the earlier work.

\section{Ellipticity Distributions}

The  relation  between the  observed  and  intrinsic distributions  of
elliptical galaxy axis ratios has  been discussed many times since the
classic paper of \cite{hubble26},  who interpreted the observations in
terms of a simple oblate model.  It is often stated that the frequency
function of apparent axis ratios cannot be used to distinguish between
different  hypotheses  for the  intrinsic  shape,  since the  observed
distribution  (a   function  of  {\it  one}   variable)  is  generally
consistent with  an infinite  set of functions  of the {\it  two} axis
ratios that define a triaxial  galaxy.  Strictly speaking, this is not
true, since an inferred distribution  may be negative for certain axis
ratios.  For   instance,  a  distribution  of   apparent  axis  ratios
$f\left(q_{\mathrm{app}}\right)\left(q_{\mathrm{app}}  \le  1 \right)$
is consistent with the oblate hypothesis only if the quantity
\begin{equation}
\int_{0}^{q_{\mathrm{app}}}\frac{f\left(x\right) dx}{\sqrt{q_{\mathrm{app}}^2 -x^2}}
\end{equation}
is  an increasing  function of  $q_{\mathrm{app}}$; violation  of this
condition  means that -- for  instance -- the  observed number  of nearly
round galaxies  is too  few to be  produced by random  orientations of
oblate spheroids, which have a tendency (stronger than that of prolate
spheroids) to appear nearly round in projection.

Until now, the positivity of the intrinsic distribution has never been
an important constraint:  deconvolutions based on the oblate, prolate
or triaxial hypotheses  have always turned out to  be non-negative for
all       axis ratios       (e.g.       \citealt{noerdlinger79,
  binggeli80,binney81}).  This  is  party  because, in  the  past,  the
distribution  of apparent ellipticities  was thought  to be  a smoothly
increasing function  of axis ratio,  with E0 galaxies the  most common
(e.g.  \citealt{sandage70}). The ellipticities  in these  studies were
usually  taken from  estimates  in  the {\it  First}  and {\it  Second
  Reference  Catalogues},  based  on  the appearance  of  galaxies  on
photographic   plates.   However,   more   recent   determinations   of
$f\left(q_{\mathrm{app}}\right)$, based on  fitting of ellipses to CCD
intensity  data for small  and homogeneous  samples ,  tend to  favor a
rather different  frequency function, with a  pronounced {\it minimum}
at  $f\left(q_{\mathrm{app}}\right)=1$. Such  a distribution  has been
found,  for  instance,  by  \citet{djorgovski86}; by  R.  Bender,  and
T. Lauer (private communications); and, in an earlier study of cluster
ellipticals,  by \citet{benacchio80}.  Figure 1  shows  a second-order
polynomial  fit to  the  \citet{djorgovski86} data,  and the  inferred
intrinsic distribution $N\left(q\right)$  under the oblate and prolate
hypothesis.  The paucity  of nearly  round galaxies  appears  to cause
problems for the oblate deconvolution: $N\left(q\right) < 0 $ for $q
> 0.93$, although  the small number  of galaxies in this  sample would
probably permit an  acceptable fit to $f\left(q_{\mathrm{app}}\right)$
with   a  positive   $N\left(q\right)$   (as  in   \citealt{franx88}).
Nevertheless the shortage of  nearly-round galaxies, if it persists in
larger samples, might eventually be shown to be inconsistent with the
oblate hypothesis. (Deconvolution algorithms like Lucy's (1974), which
{\it guarantee} positive-definite solutions, should be avoided in this
business,  since  they  may  force  a positive  solution  where  a
negative one is implied by the data.)

\begin{figure}[!htb]
\includegraphics[width=0.80\columnwidth,angle=0.]{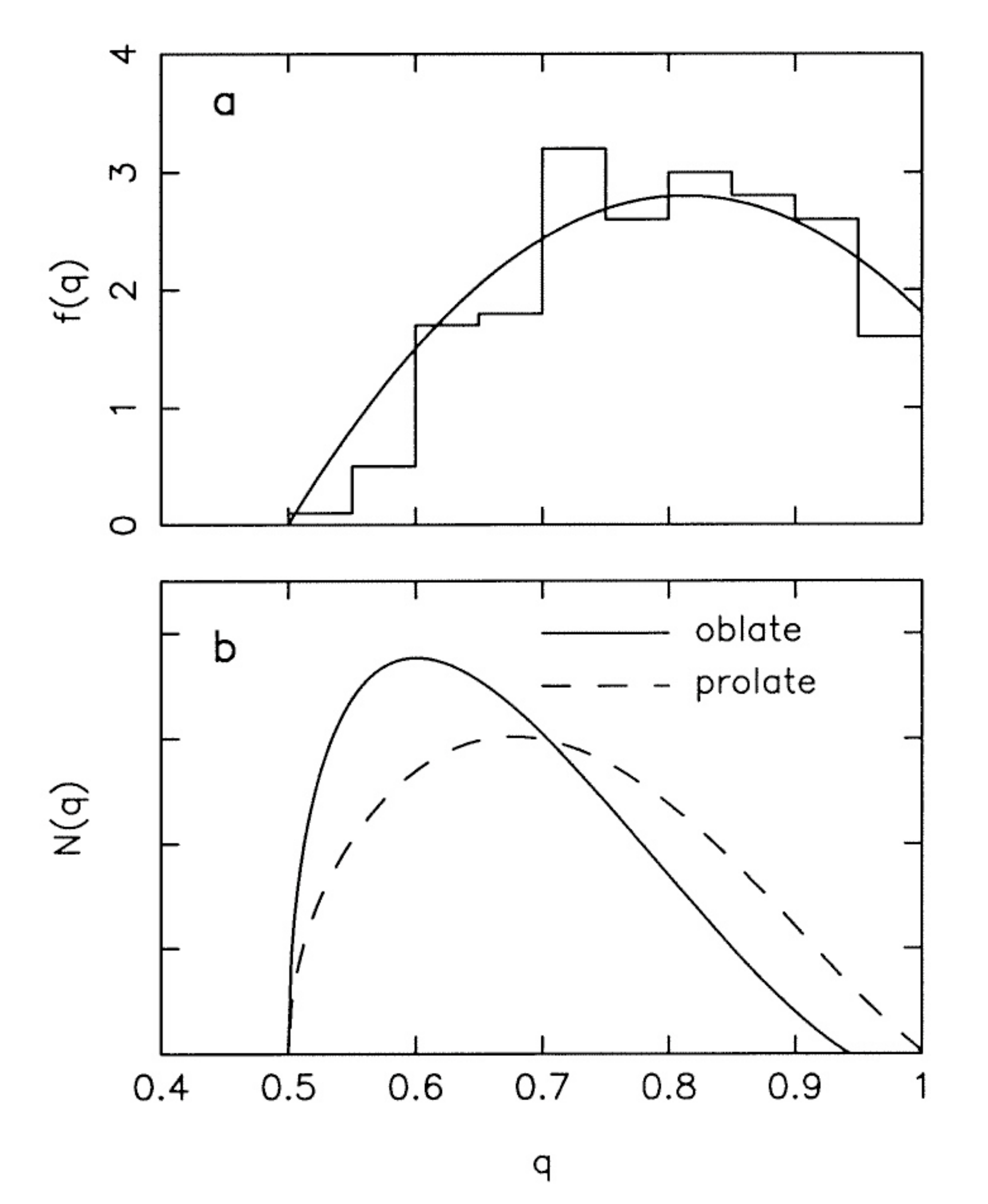}
\caption{(a)   Observed  ellipticity  distribution,   from
  \citet{djorgovski86}.  (b) Inferred  true  ellipticity distributions
  under the oblate and prolate hypotheses.}
\end{figure}

It is interesting  to speculate what sort of  galaxy formation process
would  prefer elongated galaxies  to nearly  spherical ones.  A recent
simulation  of the  formation of  halos  in the  cold dark  matter
universe  \citep{frenk88}  predicts   a  frequency  function  that  is
probably  too  strongly  weighted   toward  spherical  systems  to  be
consistent with Figure 1. Simple collapse naturally produces elongated
($\sim$  E4), prolate/triaxial  galaxies  \citep{MA:85}, but  only
from initial conditions that are unphysically smooth. Mergers of disk
systems  can easily result  in elongated  final states,  especially if
dissipation is important. Whatever  the actual formation mechanism, it
would be ironic  if the spherical galaxies so  loved by theorists were
never present in nature!

\section{Correlations with ellipticity}

The correlation  of apparent  ellipticity with surface  brightness was
first  used  as a  test  to  distinguish  between oblate  and  prolate
hypotheses by \citet{marchant79}. The surface
brightness of  an oblate  galaxy is highest  when viewed  edge-on, and
therefore when the galaxy  appears most highly flattened; the converse
is true  for a prolate galaxy.  The dependence of  the central surface
brightness $\Sigma$ on the  apparent axis ratio $q_{\mathrm{app}}$ can
be expressed simply as

\[\Sigma = \left\{ 
\begin{array}{l l}
  q_{\mathrm{app}}^{-1} \Sigma_p & \quad \mbox{(oblate)}\\
  q_{\mathrm{app}} \Sigma_p & \quad \mbox{(prolate)}\\ \end{array} \right. \]

\noindent where $\Sigma_p$ is the polar surface brightness. 
For a sample of identical   galaxies  with   random   orientations,  
the   observed correlation of  surface brightness with axis  ratio would 
immediately reveal the intrinsic shape. 
Similar  tests have been proposed based on
velocity    dispersion     \citep{lake79}    or    isophotal    radius
\citep{fasano87} instead of surface brightness.

Unfortunately,  what  works  well  for  a sample  of  identical
galaxies does not necessarily work  well for a sample of galaxies with
different axis  ratios and luminosities: even after  eliminating the
(strong) dependence  of1 surface  brightness on total  luminosity, one
has to allow  for a possible dependence of  surface brightness on {\it
  intrinsic}  ellipticity. There  is no  reason why  a  very elongated
galaxy should  have the same  polar surface brightness as  a spherical
galaxy with  the same  total luminosity. If -- for  instance -- the {\it
  equatorial}   surface    brightness   is   assumed    to   be   more
fundamental, then the relations given above should be written

\[\Sigma = \left\{ 
\begin{array}{l l}
  qq_{\mathrm{app}}^{-1} \Sigma_p & \quad \mbox{(oblate)}\\
  q^{-1}q_{\mathrm{app}} \Sigma_p & \quad \mbox{(prolate)}\\ \end{array} \right. \]

\noindent
where $q$ is the {\it intrinsic} axis ratio. For constant
$\Sigma_{eq}$ (or, rather, for a $\Sigma_{eq}$ that depends only on
total luminosity), the predicted distribution of points in the $\left(
\Sigma, q_{\mathrm{app}} \right)$ plane turns out to be almost
correlation-free.
Studies which  allow for a  dependence of polar surface  brightness or
velocity dispersion on intrinsic axis ratio \citep{merritt82,fasano89}
have so  far not produced conclusive  evidence in favor  of either the
oblate or prolate hypotheses.

Given the accumulation of  high-quality photometric and kinematic data
over the last  few years, a more sophisticated  attack on this problem
is  overdue. Probably the  correct way  to proceed  is to  assume that
elliptical galaxies define a fundamental plane something like
\begin{equation}
L \propto \Sigma_p^l \sigma_p^m q^n, 
\end{equation}
with $L$  the luminosity, $\Sigma_p$ and $\sigma_p$  the polar surface
brightness  and  velocity  dispersion,  and  $q$  the  intrinsic  axis
ratio. This  ``fundamental plane'' differs  from the one  discussed by
e.g.  \citet{djorgovski87}  in  that  (a) the  number  of  significant
variables is  increased by one;  (b) the variables defining  the plane
are all  {\it intrinsic},  and can only  be related to  {\it observed}
quantities    after   making    some   assumption    about   intrinsic
shapes. Presumably, either the  oblate or prolate hypothesis will lead
to a better  match between the predicted and  observed distribution of
galaxies  in  the space  of  observables  $\left(  L, \Sigma,  \sigma,
q_{\mathrm{app}}\right)$.  The high  dimensionality of  the  fit makes
this a tough problem, but definitely worth attempting.

\section{Isophotal Twists}

The strongest evidence  for triaxiality is
probably still the prevalence of  isophotal twists, which are a common
feature of triaxial models seen  in projection if the axis ratios vary
with radius. The  interpretation of twists in terms  of triaxiality is
not airtight: first, because galaxies may  be intrinsically twisted,
and  certainly  are  in  some   cases  of  apparent  of  recent  close
interaction; and second, because some triaxial models with ellipticity
gradients exhibit  no twists at  any viewing angle (the  best examples
being       models      with      separable       potentials;      see
\citealt{franx88}).   Nevertheless  the  abundance   of  significantly
twisted galaxies  (estimates range as  high as 50$\%$ or  60$\%$), and
the expectation  that intrinsic  twists are short-lived,  suggest that
triaxiality is common.

Little progress  has been made  on unravelling twists since  the early
papers  of \citet{carter78}, \citet{benacchio80} and \citet{leach81}. 
These  studies seemed
to show that the frequency  and amplitude of the twists was consistent
with the  observed ellipticity  gradients, assuming either  a constant
degree of  triaxiality (\citealt{benacchio80}), or a  mixture of
triaxial and axisymmetric galaxies (\citealt{leach81}). With much
larger samples,  one could imagine inferring the  distribution of axis
ratios and their  gradients $N\left(q_1, q_2, dq_1/dr,dq_2/dr \right)$
from the observed twists. Such  an analysis would be formidable, given
the fact  that both  ellipticities and position  angles often  seem to
vary in a complicated with way with radius.

Perhaps a more fundamental question -- not really addressed in the early
studies -- is  to  what degree  triaxiality  (in  the commonly  accepted
sense)    is     {\it    required}    to     explain    the    twisted
isophotes.  \citet{fasano89} have  recently attempted  to  answer this
question  by  studying  twists  in  a sample  of  isolated  elliptical
galaxies,  taking care  to  understand all  the  effects (dust  lanes,
improper flat fielding, misidentification of lenticular galaxies) that
might result in spurious twists.  They find a frequency of twisting in
their  ``unperturbed''  sample  that  is  comparable  to  the  highest
estimates given by earlier workers for more randomly selected samples;
thus  they conclude  that tidal  interactions do  not make  a dominant
contribution to  the twists.  In addition, they  find that significant
twisting is only  seen in galaxies that deviate  from a de Vaucouleurs
luminosity profile.  This they  interpret as the  signature of  a disk
superimposed on a spheroid, and they suggest that some fraction of the
twisted ellipticals may be SB0 galaxies, or else an intermediate class
of  elliptical  galaxies  with  a  disklike component.   It  would  be
interesting to determine whether there is any kinematical evidence for
this interpretation.

\section{Non-elliptical Isophotes}

Elliptical galaxies  are not precisely  ellipsoidal: deviations of  the
isophote  shapes from  perfect  ellipses, both  in  the direction  of
``boxiness''  and  ``diskiness,''  are  now routinely  measured  at  the
0.5$\%$ level  (e.g. \citealt{bender88b,michard88}). Nevertheless  it is
remarkable  that elliptical galaxies  are so  {\it well}  described by
ellipsoids.   The   first    fully   self-consistent   model   for   a
pressure-supported   triaxial   galaxy   \citep{schwarzschild79}   was
decidedly ``peanut-shaped'', and  the non-classical integrals that are
thought to  maintain triaxial figures  in the absence of  rotation are
present  in  many  models  with  strongly  non-ellipsoidal  isodensity
surfaces \citep{dezeeuw86}.  Furthermore,  the bulges of disk galaxies
are often extremely boxy (e.g. \citealt{kormendy82}).  Apparently some
physical mechanism,  as yet unguessed, guarantees  that the isodensity
surfaces  of  elliptical galaxies  never  deviate  very strongly  from
perfect ellipsoids.

To  what extent  could  the observed  deviations -- like the  isophotal
twists  discussed  above, and  the  kinematic misalignments  discussed
below -- be an artifact of  projection? Such a possibility was raised by
several     of     the     early     workers     in     this     field
(e.g. \citealt{leach81,williams81}),  but apparently never  worked out
in  quantitative detail.  The  basic idea  is that -- whereas  similar,
coaxial ellipsoids always project to perfectly elliptical isophotes --
the isophotes of a galaxy in which the ellipticity, orientation, or center
of the isodensity surfaces varies  with radius are {\it not} ellipses.
Consider  the  first  three  of   these  cases,  i.e.  a  galaxy  with
ellipticity   gradients,   but  without   any   intrinsic  twists   or
offsets.  Suppose  that the  luminosity  is  stratified on  spheroidal
surfaces of  constant $m$,  where $m^2 =  x^2 + y^2  + h\left(m\right)
z^2$. The simplest choice for $h\left(m\right)$ is $t_0 + am^2$, where
$t_0^{-1/2}$ is the  central axis ratio and $a$  is a constant related
to the ellipticity gradient; for this choice of $h\left(m\right)$, $m$
can  be expressed  simply in  terms of  the coordinates as  $m^2 =
\left(x^2  + y^2  + t_0  z^2\right) /  \left( 1  - az^2  \right)$. The
resulting model is bounded by the cylinder  $x^2 + y^2 = -t_0 / a$ for
$a <  0$, and by the  slab $z = \pm  a^{-1}$ for $a >  0$; however for
small  $a$  the bounding  surfaces  lie at  very  large  radii. For  a
power-law luminosity profile $\rho \propto  m^{-3}$ (a good fit to a de
Vaucouleurs law  near $r_e$), the  surface brightness, seen  along the
$y$ axis , is

\[\Sigma \left(x,z\right) \propto \left\{ 
\begin{array}{l l}
  \left(1-az^2\right) \sqrt{1 + ax^2/t_0}/\left(x^2 + t_0 z^2 \right), & \quad a < 0 \\
   \left(1-az^2\right)^{3/2}/\left(x^2 + t_0 z^2 \right),  & \quad a > 0.\\ \end{array} \right. \]

\begin{figure}[t]
\includegraphics[width=0.85\columnwidth,angle=0.]{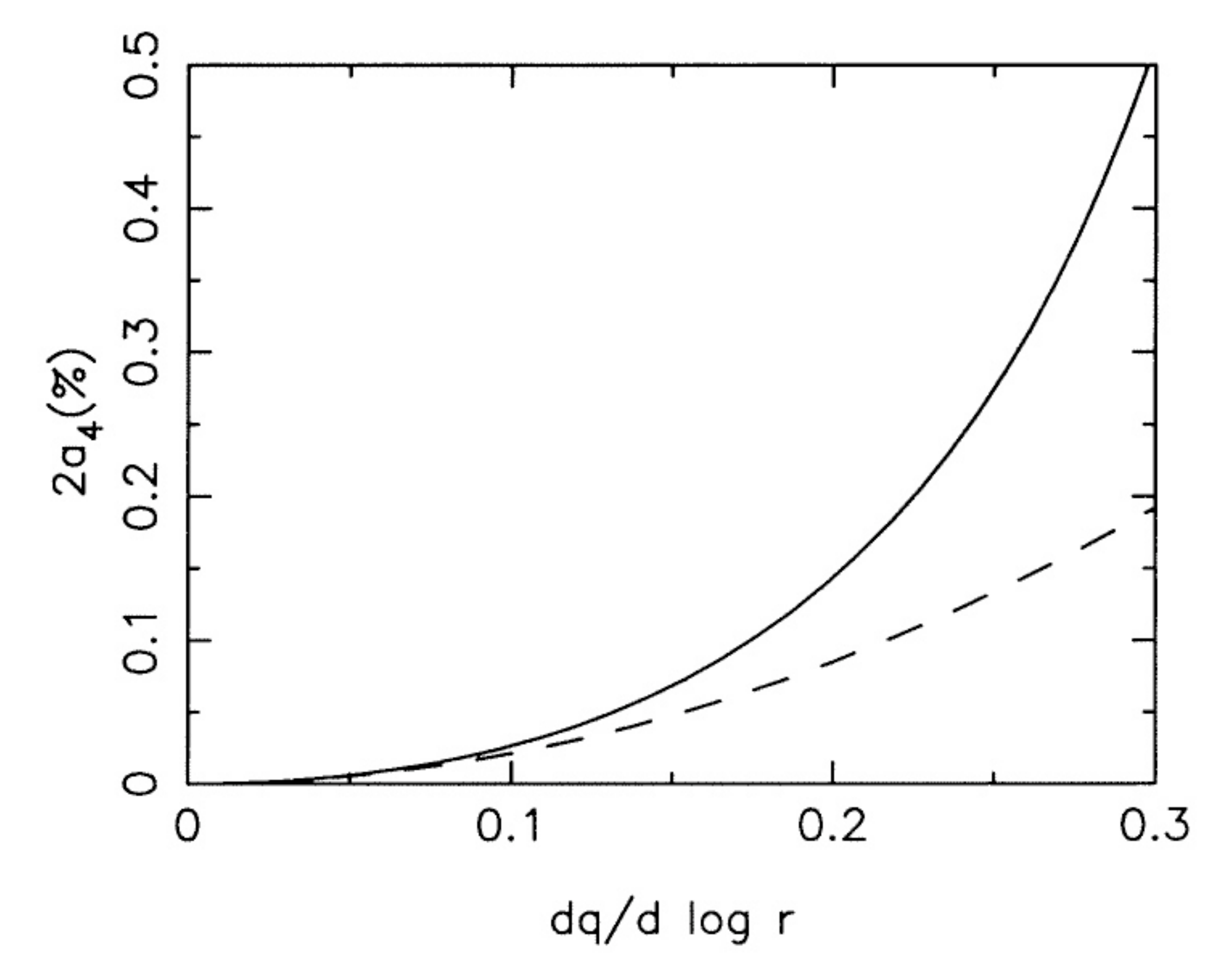}
\caption{Variation of $a_4$ with ellipticity gradient in the model discussed in the text. Solid line : major axis; dashed line : minor axis}
\end{figure}

\noindent   Define   the    ``best-fit''   ellipse   passing   through
$\left(x,0\right)$  to  be  the   ellipse  around  which  the  surface
brightness, expanded in a Fourier series
\begin{equation}
\Sigma \left(\theta \right) = \Sigma_0 \left[1+a_2 \cos(2\theta) + 2a_4 \cos(4\theta) + ...\right]
\end{equation}
in the angle $\theta =  \tan^{-1} \left(\sqrt{tz} / x \right)$, has no
$\cos(2\theta)$  component.  (The  factor  of  two  in  front  of  the
$\cos(4\theta)$ term is  due to the fact that  observers tend to quote
$a_4$ in terms  of the {\it radial} deviation  of the best-fit ellipse
from the true isophote, rather than the intensity variation around the
best-fit ellipse;  the difference is a  factor of $|d \log  \Sigma / d
\log r |  =2$ in the present  case). Setting $a_2 = 0$  fixes the axis
ratio of the best-fit ellipse passing through $\left( x, 0 \right)$:
\begin{equation}
t = t_0 + \frac{3}{2}a x^2 + ...
\end{equation}
Around this ellipse, the $\cos(4\theta)$ term has coefficient
\begin{equation}
2a_4 = \frac{3}{64}\frac{a^2 x^4}{t_0^2} + ...
\end{equation}
Note that,  to lowest order  in $ax^2$, the ellipticity  gradient just
results in  a change in  the axis ratio  of the best-fit  ellipse; the
deviations from  ``ellipticality'' are  of order $a^2x^4$,  and always
(at least in this restricted model)  go in the direction of making the
isophotes more disky ($a_4 > 0$). In terms of the ellipticity gradient
$\delta = \delta q_{\mathrm{app}} / d \log x$,
\begin{equation}
2a_4 = \frac{t_0}{48} \delta^2 + ...
\end{equation}
Figure 2 shows $a_4$ as a  function of $d q_{\mathrm{app}} / d \log r$
in  a  model with  $t_0  =  1$ (spherical  at  the  center) and  $a>0$
(increasingly  oblate).  Diskiness  at  the observed  level  evidently
requires  an ellipticity  gradient  $d q_{\mathrm{app}}  /  d \log  x$
greater  than 0.3.   \citet{bender88b} find  a few  elliptical galaxies
with $d  q_{\mathrm{app}} / d  \log x$ as  large as 0.15, but  for the
majority of galaxies in their sample, the ellipticity gradient is much
less.  Therefore  I conclude  that the observed  ellipticity gradients
fail by a factor of several  to explain the observed deviations of the
isophotes from ellipses.

Although  this  simple  analysis   appears  to  rule  out  ellipticity
gradients as an important  contributor to non-elliptical isophotes, it
would  be  worthwhile  to   extend  the  calculation  to  models  with
non-aligned or  non-concentric isodensity  surfaces. It would  also be
helpful  to understand  how one  goes about  formally  deprojecting an
observed  intensity  distribution   to  obtain  the  three-dimensional
luminosity  distribution, {\it without}  assuming that  the isodensity
surfaces are ellipsoids, and without making an {\it ad hoc} assumption
(as   above)    about   the    radial   dependence   of    the   shape
parameters. \citet{rybicki87} has carried out some preliminary work in
this  direction;  the problem  is  inherently  underdetermined in  the
absence of information about the inclination angle.

The next simplest interpretation  of the non-elliptical isophotes is a
model in which the  underlying elliptical galaxy is, again, accurately
ellipsoidal,  but  with   a  second  component  (presumably  different
kinematically)  that  has   a  different  morphology.  ``Disky''
isophotes are  naturally modeled by superimposing  an exponential disk
on           an           otherwise          normal           spheroid
(e.g. \citealt{carter87,jedrzejewski87,rix90}).  Disky ellipticals are
often  found to  be rotationally  supported \citep{bender88a},  and many
such     galaxies     may     simply     be     misclassified     S0's
(e.g. \citealt{capaccioli88}). No  comparably simple interpretation of
``boxy'' isophotes has  so far been suggested. Two  conclusions may be
drawn  from  these studies:  first,  that  many elliptical  galaxies
probably contain a  more or less significant disk;  and second; to the
extent  that  the  non-elliptical  isophotes are  due  to  photometric
superposition  of a  spheroid and  a disk,  the  underlying elliptical
galaxies  must be  even  {\it more}  accurately  ellipsoidal than  the
measured values of $a_4$ would suggest.

\section{Kinematic Tests}

These are of two basic kinds. The rings of dust and
gas seen in a few elliptical galaxies can sometimes be identified with
periodic orbits.   Given an  assumption about the  radial form  of the
underlying potential  (which may  or may not  trace the light),  it is
then possible to  infer the intrinsic shape of  the potential from the
behavior of the orbit. F. Bertola and T. de Zeeuw discuss this sort of
analysis in their contributions to these proceedings.

The  second  kind  of kinematical  test  is  based  on a
  geometrical  property  of  triaxial  ellipsoids. Because  the  
  apparent minor  axis of  a triaxial galaxy  need not  be coincident
with the projection of the intrinsic minor axis, it follows that
stellar  streaming  around the  intrinsic  minor  axis will  generally
result in some component  of line-of-sight velocity along the apparent
minor  axis \citep{binney85}.  Thus  the observation  of minor  axis
rotation  provides a  certain amount  of  evidence in  favor of  the
triaxial hypothesis.  As always, the inference is  not airtight, since
(a) stellar streaming may occur around the intrinsic {\it major} axis;
i.e.  a prolate  ``spindle''; (b)  elliptical galaxies  certainly have
rotating figures,  and the rotation axis  may even be  inclined to the
symmetry  axis.  However  observations  show a  preference  for  small
kinematic   misalignments   (e.g.   \citealt{davies88}),   and   large
misalignments are often seen in galaxies with large photometric twists
\citep{franx89}, suggesting that triaxiality is the main culprit.

As in the  case of isophotal twists, making  detailed inferences about
the shapes of galaxies from  statistics of the observed rotation fields
is  extremely difficult.   Since  a  major study  of  this problem  is
nearing completion \citep{franx90}, I  will limit myself to an outline
of the main points. In its  simplest form, the problem is very similar
to the surface  brightness / ellipticity test described  above: given
some measure  of the kinematic  misalignment for a sample  of galaxies
(e.g., the  ratio of minor to  major axis rotation  velocities on some
isophote), what hypothesis for the distribution of intrinsic shapes is
most consistent  with the observed correlation of  this parameter with
apparent  axis  ratio,  assuming  that  the  streaming  is  about  the
intrinsic minor  axis? \citet{franx88} finds  that a triaxiality  $Z =
\left( 1 -  q_2 \right) / \left(1 - q_1 \right)  \approx 0.5$ fits the
data well,  except for a  possible abundance peak at  misalignments of
$\sim  90^\circ$, indicative  of rotation  around the  intrinsic major
axis.

As  Franx points  out,  a major  uncertainty  in his  analysis is  the
likelihood of {\it intrinsic}  misalignments: since triaxial galaxies
contain families  of tube orbits  that circulate both about  the major
and minor  axes, there  is no good  reason to expect  streaming around
only one  of these axes.  Furthermore, the intrinsic  misalignment may
itself be correlated  with intrinsic axis ratios, in  a way that masks
the the  expected correlation between observed quantities  (just as in
the surface  brightness / ellipticity  test discussed above).  Work in
progress  should elucidate to  what extent  these difficulties  can be
overcome.

\section{Stability}

One of the reasons that the oblate hypothesis persisted so long in the
minds of galactic dynamicists is  that the number of orbital integrals
in axisymmetric  potentials (always  two, sometimes three)  is clearly
sufficient  to allow the  existence of  self-consistent models  of any
axis ratio. Schwarzschild's (1979) construction of an equilibrium {\it
  triaxial} model  put an end  to this mind  set: clearly it  was now
impossible to prefer axisymmetric  over non-axisymmetric models on the
basis  of equilibrium  arguments alone.  \citet{statler87} went  on to
demonstrate that essentially  no point in the plane  of the model axis
ratios could be  excluded by the requirement of  equilibrium, at least
for models based on the  so-called ``perfect'' potentials in which all
parts   of  phase   space   are  characterized   by  three   isolating
integrals.  But  it has  recently  become  clear  that many  of  these
equilibrium models are dynamically unstable.

Most  of the  work to  date on  stability of  non-rotating  models has
focused on the spherical  case. While isotropic
spherical models are  generally stable \citep{antonov62}, surprisingly
small  amounts  of radial  anisotropy  can  induce  instability to  an
$m=2$  or  ``bar'' mode,  which  causes the  model  to  evolve into  a
triaxial spheroid. A recent study of this ``radial-orbit instability'' in a
particular family  of spherical isochrone  models \citep{saha90} finds
instability  for $\sigma_r  / \sigma_t  > 1.2$,  where  $\sigma_r$ and
$\sigma_t$ are the mean radial and (one-component) tangential velocity
dispersions. Work  in progress should soon reveal  whether this result
holds for a  wider class of anisotropic spherical  models. Although it
would  probably be  extreme to  blame this  instability alone  for the
apparent absence of nearly-spherical galaxies (Figure 1), it is likely
that  radial  collapse sometimes  produces  bar-unstable final  states
\citep{MA:85}, thus leading to a preference for nonspherical (even
non-axisymmetric) galaxies over spherical (axisymmetric) ones.

\begin{figure}[t] 
\includegraphics[width=1.0\columnwidth,angle=0.]{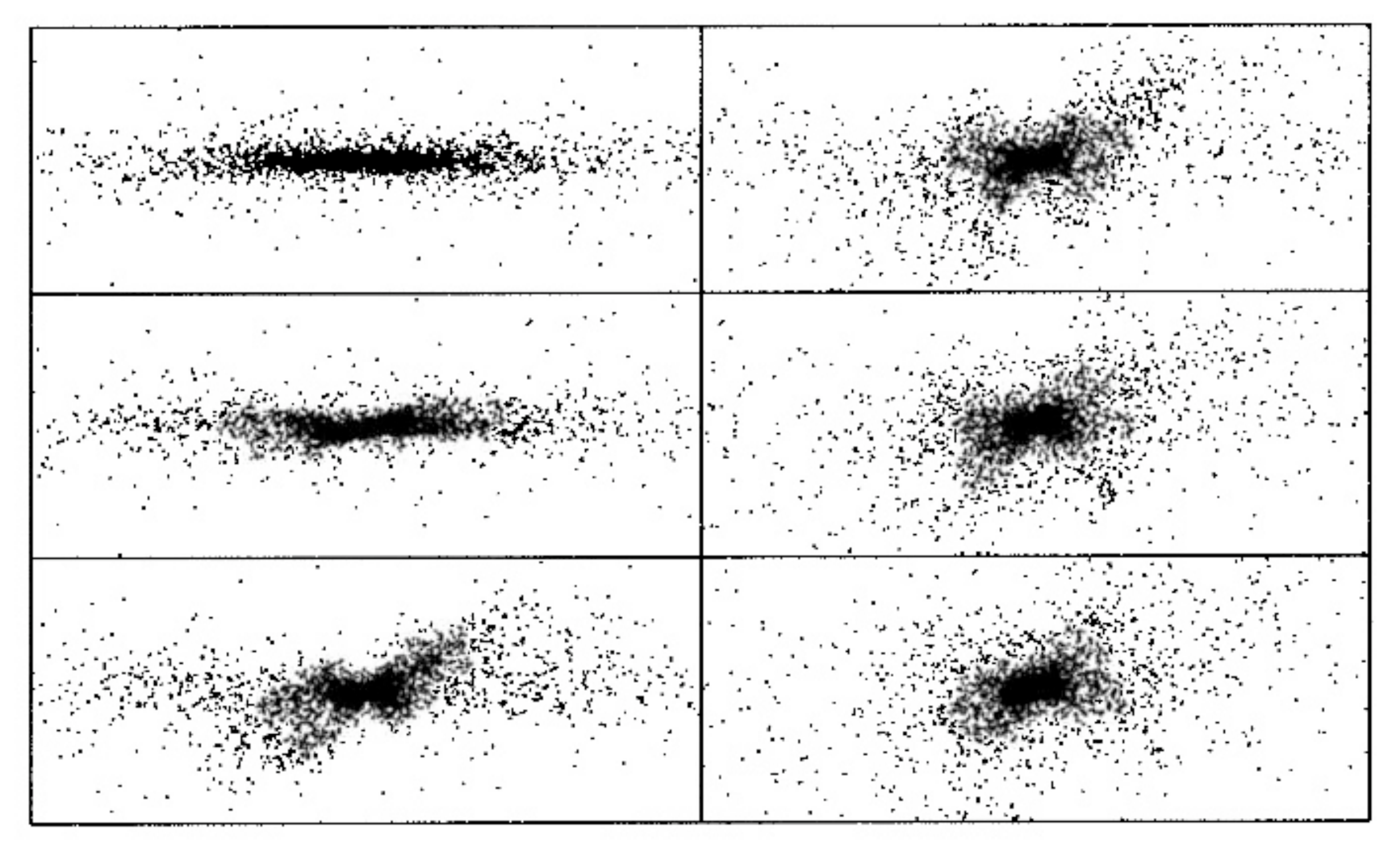}
\caption{Evolution  of a prolate model with
initial axis ratio 1 : 10 (from \citealt{merritt91}).}  
\end{figure}

At the other extreme of the ellipticity distribution, it appears more and
more likely that the absence of elliptical galaxies flatter than about
E6 can be ascribed  to instabilities. \citet{fridman84} suggested that
oblate and prolate models with axis ratios more extreme than about 2 :
5 might be generally unstable to bending models, a prediction that has
now    been   verified    in    one   family    of   prolate    models
\citep{merritt91}.  Figure  3  shows  the  instability  acting  on  an
initially E9  model.  Demonstrating the existence  of this instability
in a range of prolate and oblate models will require considerably more
work, but may solve once and for all the classic problem of why highly
flattened elliptical galaxies to not exist.

\clearpage

\bibliography{ms}

\begin{thebibliography}{39}
\expandafter\ifx\csname natexlab\endcsname\relax\def\natexlab#1{#1}\fi

\bibitem[{{Antonov}(1962)}]{antonov62}
{Antonov}, V.~A. 1962, Vestnik Leningrad Univ., 19, 96

\bibitem[{{Benacchio} \& {Galletta}(1980)}]{benacchio80}
{Benacchio}, L., \& {Galletta}, G. 1980, \mnras, 193, 885

\bibitem[{{Bender}(1988)}]{bender88a}
{Bender}, R. 1988, \aap, 193, L7

\bibitem[{{Bender} {et~al.}(1988){Bender}, {Doebereiner}, \&
  {Moellenhoff}}]{bender88b}
{Bender}, R., {Doebereiner}, S., \& {Moellenhoff}, C. 1988, \aaps, 74, 385

\bibitem[{{Binggeli}(1980)}]{binggeli80}
{Binggeli}, B. 1980, \aap, 82, 289

\bibitem[{{Binney}(1985)}]{binney85}
{Binney}, J. 1985, \mnras, 212, 767

\bibitem[{{Binney} \& {de Vaucouleurs}(1981)}]{binney81}
{Binney}, J., \& {de Vaucouleurs}, G. 1981, \mnras, 194, 679

\bibitem[{{Capaccioli} {et~al.}(1988){Capaccioli}, {Piotto}, \&
  {Rampazzo}}]{capaccioli88}
{Capaccioli}, M., {Piotto}, G., \& {Rampazzo}, R. 1988, \aj, 96, 487

\bibitem[{{Carter}(1978)}]{carter78}
{Carter}, D. 1978, \mnras, 182, 797

\bibitem[{{Carter}(1987)}]{carter87}
---. 1987, \apj, 312, 514

\bibitem[{{Davies} \& {Birkinshaw}(1988)}]{davies88}
{Davies}, R.~L., \& {Birkinshaw}, M. 1988, \apjs, 68, 409

\bibitem[{{de Zeeuw} {et~al.}(1986){de Zeeuw}, {Peletier}, \&
  {Franx}}]{dezeeuw86}
{de Zeeuw}, T., {Peletier}, R., \& {Franx}, M. 1986, \mnras, 221, 1001

\bibitem[{{Djorgovski} \& {Davis}(1987)}]{djorgovski87}
{Djorgovski}, S., \& {Davis}, M. 1987, \apj, 313, 59

\bibitem[{{Djorgovski}(1986)}]{djorgovski86}
{Djorgovski}, S.~B. 1986, Diss.~Abstr.~Int., Sect.~B, Vol.~47, No.~3, p.~1102,
  47, 1102

\bibitem[{{Fasano}(1987)}]{fasano87}
{Fasano}, G. 1987, in IAU Symposium, Vol. 127, Structure and Dynamics of
  Elliptical Galaxies, ed. P.~T. {de Zeeuw}, 395

\bibitem[{{Fasano} \& {Bonoli}(1989)}]{fasano89}
{Fasano}, G., \& {Bonoli}, C. 1989, \aaps, 79, 291

\bibitem[{{Franx}(1988)}]{franx88}
{Franx}, M. 1988, PhD thesis, University of Leiden, The Netherlands

\bibitem[{{Franx} {et~al.}(1991){Franx}, {Illingworth}, \& {de
  Zeeuw}}]{franx90}
{Franx}, M., {Illingworth}, G., \& {de Zeeuw}, T. 1991, \apj, 383, 112

\bibitem[{{Franx} {et~al.}(1989){Franx}, {Illingworth}, \& {Heckman}}]{franx89}
{Franx}, M., {Illingworth}, G., \& {Heckman}, T. 1989, \apj, 344, 613

\bibitem[{{Frenk} {et~al.}(1988){Frenk}, {White}, {Davis}, \&
  {Efstathiou}}]{frenk88}
{Frenk}, C.~S., {White}, S.~D.~M., {Davis}, M., \& {Efstathiou}, G. 1988, \apj,
  327, 507

\bibitem[{{Fridman} \& {Polyachenko}(1984)}]{fridman84}
{Fridman}, A.~M., \& {Polyachenko}, V.~L. 1984, {Physics of gravitating
  systems} (New York: Springer, 1984)

\bibitem[{{Hubble}(1926)}]{hubble26}
{Hubble}, E.~P. 1926, \apj, 64, 321

\bibitem[{{Jedrzejewski} {et~al.}(1987){Jedrzejewski}, {Davies}, \&
  {Illingworth}}]{jedrzejewski87}
{Jedrzejewski}, R.~I., {Davies}, R.~L., \& {Illingworth}, G.~D. 1987, \aj, 94,
  1508

\bibitem[{{Kormendy} \& {Illingworth}(1982)}]{kormendy82}
{Kormendy}, J., \& {Illingworth}, G. 1982, \apj, 256, 460

\bibitem[{{Lake}(1979)}]{lake79}
{Lake}, G. 1979, in Photometry, Kinematics and Dynamics of Galaxies, ed. D.~S.
  {Evans}, 381

\bibitem[{{Leach}(1981)}]{leach81}
{Leach}, R. 1981, \apj, 248, 485

\bibitem[{{Marchant} \& {Olson}(1979)}]{marchant79}
{Marchant}, A.~B., \& {Olson}, D.~W. 1979, \apjl, 230, L157

\bibitem[{{Merritt}(1982)}]{merritt82}
{Merritt}, D. 1982, \aj, 87, 1279

\bibitem[{{Merritt} \& {Aguilar}(1985)}]{MA:85}
{Merritt}, D., \& {Aguilar}, L.~A. 1985, \mnras, 217, 787

\bibitem[{{Merritt} \& {Hernquist}(1991)}]{merritt91}
{Merritt}, D., \& {Hernquist}, L. 1991, \apj, 376, 439

\bibitem[{{Michard} \& {Simien}(1988)}]{michard88}
{Michard}, R., \& {Simien}, F. 1988, \aaps, 74, 25

\bibitem[{{Noerdlinger}(1979)}]{noerdlinger79}
{Noerdlinger}, P.~D. 1979, \apj, 234, 802

\bibitem[{{Rix} \& {White}(1990)}]{rix90}
{Rix}, H.-W., \& {White}, S.~D.~M. 1990, \apj, 362, 52

\bibitem[{{Rybicki}(1987)}]{rybicki87}
{Rybicki}, G.~B. 1987, in IAU Symposium, Vol. 127, Structure and Dynamics of
  Elliptical Galaxies, ed. P.~T. {de Zeeuw}, 397

\bibitem[{{Saha}(1991)}]{saha90}
{Saha}, P. 1991, \mnras, 248, 494

\bibitem[{{Sandage} {et~al.}(1970){Sandage}, {Freeman}, \&
  {Stokes}}]{sandage70}
{Sandage}, A., {Freeman}, K.~C., \& {Stokes}, N.~R. 1970, \apj, 160, 831

\bibitem[{{Schwarzschild}(1979)}]{schwarzschild79}
{Schwarzschild}, M. 1979, \apj, 232, 236

\bibitem[{{Statler}(1987)}]{statler87}
{Statler}, T.~S. 1987, \apj, 321, 113

\bibitem[{{Williams}(1981)}]{williams81}
{Williams}, T.~B. 1981, \apj, 244, 458

\end{thebibliography}

\end{document}